\documentclass[12pt,preprint]{aastex}

\shorttitle{X-ray Light Curve of V1494 Aql}
\shortauthors{Drake et al.}

\begin{document}

\title{The Extraordinary X-ray Light Curve of the Classical Nova
V1494 Aquilae (1999 \#2) in Outburst: The Discovery of Pulsations
and a ``Burst''}



\author{Jeremy J.~Drake\altaffilmark{1}, R. Mark Wagner\altaffilmark{2}, 
Sumner Starrfield\altaffilmark{3}, Yousaf Butt\altaffilmark{1}, 
Joachim Krautter\altaffilmark{4}, H. E. Bond\altaffilmark{5}, M. 
Della Valle\altaffilmark{6}, R. D. Gehrz\altaffilmark{7}, Charles E. 
Woodward\altaffilmark{7}, A. Evans\altaffilmark{8}, M. 
Orio\altaffilmark{9}, P. Hauschildt\altaffilmark{10}, M. 
Hernanz\altaffilmark{11}, K. Mukai\altaffilmark{12}, \& J. W. 
Truran\altaffilmark{13}}

\altaffiltext{1}{Smithsonian Astrophysical Observatory, 60 
Garden Street, Cambridge, MA 02138; (jdrake,ybutt)@cfa.harvard.edu}
\altaffiltext{2}{Large Binocular Telescope Observatory, 933 North Cherry Avenue, 
Tucson, Arizona 85721; rmw@as.arizona.edu}
\altaffiltext{3}{Department of Physics \& Astronomy, Arizona State 
University, Tempe, Arizona 85287; sumner.starrfield@asu.edu}
\altaffiltext{4}{Heidelberg-Koenigstuhl Landessternwarte, Heidelberg 
1, D-69121 Germany}
\altaffiltext{5}{Space Telescope Science Institute, 3700 San Martin 
Drive, Baltimore, MD 21218; bond@stsci.edu}
\altaffiltext{6}{Dipartamento di Astronomia e Scienza dello Spazio e 
Osservatorio Astronomico di Arcetri, Largo Enrico Fermi, 5, Firenzi, 
I-50125 Italy}
\altaffiltext{7}{Department of Astronomy, University of Minnesota, 
116 Church Street SE, Minneapolis, MN 55455; 
(gehrz,chelsea)@astro.umn.edu}
\altaffiltext{8}{Department of Physics, Keele University, Keele, 
Staffordshire, ST5 5BG}
\altaffiltext{9}{Oss. Astronomico di Torino, Strada Osservatorio 20, 
Pino Torinese I-10025 Italy}
\altaffiltext{10}{Hamburger Sternwarte, Gojenbergsweg 112, 
21029 Hamburg, Germany; phauschildt@hs.uni-hamburg.de}
\altaffiltext{11}{Instituto de Ciencias del Espacio (CSIC), \& 
Institut d'Estudis Espacials de Catalunya, Edifici Nexus-201, C/ Gran 
Capit\`{a} 2-4, E-08034 Barcelona, Spain}
\altaffiltext{12}{NASA Goddard Space Flight Center, Code 662, 
Laboratory for High Energy Astrophysics, Greenbelt, MD 20771; 
mukai@milkyway.gsfc.nasa.gov}
\altaffiltext{13}{Department of Astronomy \& Astrophysics, University 
of Chicago, 5640 Ellis Avenue, Chicago, IL 60637; 
truran@nova.uchicago.edu}

\begin{abstract}
V1494 Aql (Nova Aql 1999 No. 2) was discovered on 2
December  1999.  We obtained {\it Chandra} ACIS-I spectra on 15 April
and 7 June 
2000 which appear to show only emission lines.  Our third
observation, on 6 August, showed that its spectrum had evolved
to that characteristic of a Super Soft X-ray Source. 
We then obtained {\it Chandra} LETG+HRC-S spectra on 28
September (8 ksec) and 1 October (17  ksec).
We analyzed the X-ray light curve of our grating
observations and found both a short time scale ``burst'' and
oscillations.  Neither of these phenomena have previously been seen in
the light curve of a nova in outburst.  The ``burst'' was a factor
of $\sim$10 rise in X-ray counts near the middle of the second
observation, and which lasted about 1000 sec; it exhibited at least
two peaks, in addition to other structure.  Our time series analysis of the
combined 25 ksec observation shows a peak at $\sim$2500~s which
is present in independent analyses of both the zeroth order
image and the dispersed spectrum and is not present in similar
analyses of grating data for HZ 43 and Sirius B.
Further analyses of the V1494 Aql data find other periods
present which implies that we are observing non-radial g$^+$ modes
from the pulsating, rekindled white dwarf.
\end{abstract}



\section{Introduction}

Classical novae (CN) are the third most violent stellar
explosions after $\gamma$-ray bursts and supernovae.   At
maximum light, their bolometric luminosities exceed
$10^5$L$_\odot$, and they eject more than
$10^{-4}$M$_\odot$ of material into space, with strongly
non-solar compositions.
Previous X-ray (XR) studies of novae {\it in outburst} were
performed by EXOSAT, ROSAT, and most recently ASCA,
{\it Bepposax} and {\it Chandra}.  EXOSAT detected QU Vul, PW
Vul, RS Oph, and GQ Mus in the constant bolometric
luminosity phase (\"Ogelman et al. 1987).  ROSAT detected
V838 Her, V351 Pup, V1974 Cyg, GQ Mus, and LMC 1995
(Krautter et al. 1996, K96; Orio et al. 2001a). {\it The analyses of these
data show that no other wavelength region provides
unambiguous information on the evolution of the thermonuclear runaway on the
white dwarf} (WD).  The ROSAT studies identified two components of the
XR emission from a nova in outburst.  The first is a soft
component that places novae at, or near, the bright end of the
class of Super Soft X-ray Sources (hereafter SSS; at low
resolution the spectrum resembles an ``emission'' line with a peak $\sim$
0.5 keV).  There is also a ``hard'' component, $\sim$ 1 keV, that was
seen in 3 of the 5 novae studied with ROSAT (K96, Orio et al. 2001a;
and in V382 Vel studied
with {\it Bepposax} 
and {\it Chandra} (Orio et al. 2001a; Starrfield et al. 2000).
It was the ROSAT studies of V1974 Cyg, however, that established
the connection between the SSS and CN in outburst (K96).
It stayed bright in XR for more than
18 months and reached a peak of 76 cts s$^{-1}$ before
beginning its decline to quiescence (K96; Balman et al. 1998,
B98).

V1494 Aquilae (Nova Aql 1999 \#2) was discovered in the
optical at m$_{v}\sim 6$ on 1999 December 1.785 UT by
Pereira (1999). 
This nova then reached a peak magnitude
exceeding 4 within a few days of discovery.
A report on early optical spectroscopic
observations can be found in Kiss \& Thomson (2000) who
described it as a member of the Fe II class as defined by
Williams (1992).  This suggests that the nova ejected an
optically thick cloud of material.

We activated our {\it Chandra} Target - of Opportunity (ToO)
Program in January 2000 and obtained three observations with
the ACIS - I instrument followed by two observations with the
LETG+HRC-S instrument (separated by 4 days because of a
$\gamma$-ray burst ToO).  Our first ACIS-I spectrum was
obtained on 15 April 2000.  Both it and a second ACIS-I
spectrum (7 June) showed an emission line spectrum with no
obvious soft component (Starrfield et al. 2000; 2001).  Our third
ACIS-I spectrum on 6 August, however, showed that the nova
had evolved into a bright SSS 
with a few weak emission lines at higher energies
(Starrfield et al. 2000).  We obtained an LETG+HRC-S
spectrum at the end of September and this observation showed
the high dispersion spectrum of a SSS.  While it resembles the
XMM spectrum of the SSS CAL 83 (Paerels et al. 2001), it
appears somewhat hotter and the ``emission features'' are at
different wavelengths. The evolution from a hard source to that
of a SSS was also seen in the ROSAT observations of V1974
Cyg (K96) and combined ASCA-Bepposax observations of
V382 Vul (Orio et al. 2001b, 2002; Mukai \& Ishida 2001).

We will report on the spectral development and analyses of the
spectra elsewhere. Because of its spectacular nature, in this 
paper we concentrate only on the light
curve of the XR data obtained with {\it Chandra}.  In the next
section we describe the observations and the method of
obtaining the light curve.  In section 3 we describe
our time-series analyses, section 4 contains a discussion of
our results, and the last section is the
summary.

\section{Observations and Analysis}

V1494 Aql was observed in two segments using the
LETG+HRC-S.  The first observation (ID 2308) was carried out
on 2000 September 28 between UT~06:51:13 and 09:36:56, and
the second (ID 72) three days later on 2000 October 1, between
UT~10:08:56 and 15:39:30.  Initial reduction of satellite
telemetry was performed by the {\it Chandra} X-ray Center
Standard Data Processing software version R4CU5UPD9.  The
analysis described here is based on the Level 2 products of this
processing, together with instrument dead time corrections that
are products of Level 1 processing.  Dead time corrections
generally need to be taken into account in the determination of
exposure times when the total detector count rate in a particular
interval exceeds the telemetry limit of 183 count~s$^{-1}$. The
computed corrections are a multiplicative factor that, when
averaged and applied to the elapsed exposure time for a
particular time interval, yield the effective exposure time for that
interval.  Both of our observations experienced episodes of
telemetry saturation and we have included dead time effects in
all the subsequent analyses.  However, these episodes were
sufficiently short and infrequent so as to have no significant
affect on the derived light curves.  Total effective exposure
times for the first and second observation were 8146 and
18232s, respectively.

We computed light curves from the bright zeroth order of the
LETG+HRC-S spectrum as seen in the Level 2 photon event list
using the PERL program LCURVE (P.~Ratzlaff, unpublished).
A circular source region with a diameter of 10 arcsec was
employed, together with a surrounding annular region for
background estimation, sized so as to be a factor ten larger in
surface area than the source region.  The source region was
sufficiently large so as to include all significant signal in the
LETG coarse support structure diffraction wings, and the
background region sufficiently small so as not to include any
significant amount of the first order dispersed spectrum.

\section{Results}

\subsection{Light Curves}

Our light curves are illustrated in Figures~1 and 2.  The first of
these shows the entire observations of first (ID 72) and second (ID
2308) segments using 25~s temporal bins.  Note that all three plots
cover the same size time interval (20~ks).  The upper two panels of
this figure cover the same range in count rate, while the lower panel
shows the full range of count rate variation scaled by the flare
observed near the middle of the second segment.  Readily apparent in
both segments is stochastic variability on time scales of minutes.  In
addition, a slower modulation with a period $\sim$2000-3000~s appears
to be present. In the third panel, the flare corresponds to a
$\sim$10-fold increase in intensity over the quiescent background
signal and lasts for $\sim$820~s.

In Figure~2, we show the region around the time of the outburst in the
second segment with a temporal bin size of 25~s.  It is clear that the
flare is not a single isolated event, but exhibits two main large
flares and perhaps a precursor and trailer.  Examining Figure~2, a
precursor to the main flare appears to begin at t $\simeq$ 8850 with a
3-fold increase in intensity and lasts for $\sim$270~s.  This event is
rapidly followed by an increase of $\sim$2.5 over the next 130~s to a
well-defind narrow peak which lasts $\sim$60~s.  A local minimum is
reached at t $\simeq$ 9340 but the intensity increases rapidly again
peaking at t $\simeq$ 9390, a factor of $\sim$10 above the quiescent
background intensity.  The duration of the second flare appears
comparable to the first.  Approximately 110~s later, the intensity has
declined to nearly the quiescent level, but appears to be quickly
followed by a small trailing flare event which lasts for $\sim$170~s.
By t $\simeq$ 9670, the burst is over.

\subsection{Timing Analysis}

In order to better understand the nature of the temporal variations
seen in the light curves and discussed above, we performed four
different time series analyses, each with increasing sophistication.
These included: (1) a simple periodogram formed by fitting a cosine
function over an array of trial periods; (2) phase dispersion
minimization (PDM: Stellingwerf 1978);  (3) a Fourier power
spectrum (Deeming 1975); and (4) a "cleaned" Fourier power spectrum (Roberts,
et al. 1987) in which the sampling or spectral window function is
removed from the raw or dirty power spectrum.  This latter robust algorithm
is particularly well suited to the analysis of a multiperiodic time
series of unequally spaced data.

In preparation for the 4 techniques discussed above, the flare
detected in the second segment was excised from the 
HRC-S photon event lists.  This removal created three
separate segments, each with a duration of 8000-9000~s: a segment
consisting of the entire first observation (ID 2308) followed by two
segments from the second observation (ID 72).  These data are sampled
in 25~s temporal bins, and are sensitive to periodic modulations with
periods ranging from twice the bin width (50 s) to one half of the
total time span covered by the observations (13,000 s).  

We found that the time series spectra derived from a simple cosine
periodogram, a PDM, and a Deeming power spectrum, are characterised by
a significant periodicity at 2500~s as well as other possibly
significant features.
This interpretation is complicated by the fact that strong sidelobes
of the sampling or window function produce spurious features
throughout the power spectrum and thus affect the shape and power of
significant features.  Many attempts have been made to recognize these
spurious features (Deeming 1975) but they remain problematic.

Recently, Roberts et al. (1987) have developed a one--dimensional
variation of the two dimensional {\sc{CLEAN}} (H\"{o}gbom 1974)
algorithm which is well suited to the temporal analysis of unequally
spaced finite data samples as we consider here.  We have applied this
version of the {\sc CLEAN} algorithm to our X-ray time series data of
V1494 Aql, binned this time at 75s intervals, 
and the results are shown in Figure~3.  Using this algorithm,
the spectral window function has been iteratively subtracted in the
Fourier domain from the raw or dirty power spectrum shown in
Figure~3 (top panel).  We assumed a gain of 0.2 and stopped after 10,000
iterations since no further improvement in the residual spectrum was
apparent.  Small gain values prevent the accumulation of errors in the
subtraction of the sidelobes at the expense of computation time, and
are preferred over gains close to unity and fewer iterations.  The
data were averaged in 75~s bins for this analysis.  A frequency
resolution of 1.0 $\times$ 10$^{-5}$ Hz was chosen.

The cleaned power spectrum is shown in the bottom panel of Figure~3.
It is difficult to attach an overall significance level to features
appearing in the power spectrum because it depends on both the
sampling and the signal-to-noise ratio of the data.  To help assess
the reality of features appearing in the power spectrum, therefore, we
added a tracer periodic signal to the time series data with a
frequency of 0.0011 Hz and an amplitude of 5\% of the mean signal
level.  The location of this signal is indicated in Figure~3 and we
note that the signal is easily detected suggesting that significant
signals in this frequency range might have amplitudes of a few
percent.  The strongest peak is at $\nu$ = 0.0004002 Hz (P = 2498.8~s)
and has an amplitude of 15\% (0.11 counts/s) of the mean count rate
level.  The strength and detectability of the artificial signal
suggests that other comparably strong features are real as well and
thus the X-ray variations might be multiperiodic.  These include a
double--peaked feature at 0.00029 Hz (3461~s), 0.00059 Hz (1695~s),
0.00079 Hz (1266~s), 0.00097 Hz (1031~s), 0.00124 Hz (809~s), and
perhaps 0.0019 Hz (526~s).

In addition to our main analysis of the 75~s binned data, we also
examined the cleaned power spectra of these data averaged in 25~s and
150~s temporal bins to further assess their reality.  The features
listed above all appear in these additional power spectra at
significant levels as defined by the tracer signal that we
introduced into the data.   

After further investigation, we found that the 2500~s period is close
to the beat period of 2623.98~s formed from the two orthogonal
spacecraft dither motions (with periods of 768.57~s and 1086.96~s in
pitch and yaw, respectively).  However, 
neither the pitch or yaw periods are present in the power spectra so
it is difficult to understand how the beat period could be present in
isolation.  Nevertheless, in order to determine whether or not the
2500~s periodicity might be related to the spacecraft motion we
performed identical analyses on light curves derived from calibration
observations of the hot white dwarf stars HZ~43 and Sirius~B (ObsID's
59 and 1452).  These light curves were expected to be featureless, and
indeed their periodograms were mostly flat over the interval from 300
to 4000~s.  The power spectrum of Sirius B shows a broad ($\Delta$t
$\simeq$ 500~s FWZI) and a weak feature at 2500~s but it is significant
at less than the 2$\sigma$ level and thus not characteristic of the
precise spacecraft dither motions.

We also used the {\it Chandra}
Interactive Analysis of Observations (CIAO) software to isolate the
regions around the dispersed spectra and
obtained the light curve for these regions which were sampled in 50~s
bins.  We then performed the same cleaned power spectrum analyses on
the light curve of the dispersed spectra.  This analysis thus serves
as an independent check on the results described above.  We find the
same major period at $\sim$2500~s is present in the dispersed light as
well as the additional periods listed above.  These results, are {\it
obtained completely independently of the zeroth order light curve},
and strengthen our conclusion that the observed periods are not
instrumental.  

We can safely exlude instrumental background as being responsible for
any or part of the time-variable signals we have detected.  In the
case of the 0th order light curve, for example, the background rate
scaled to the source region size amounts to 0.5-$1\,
10^{-2}$~count~s$^{-1}$, as compared to the quiescent source rate of
$\sim 0.6$~count~s$^{-1}$: the background is completely negligible.
We also examined the Chandra on-board particle monitor and found that
no unusual episodes occurred during the interval of our V1494~Aql
observation.

Another interesting feature is present in the cleaned power spectrum.
A strong signal appears at a frequency of $9.03 \times 10^{-5}$ Hz
which is consistent with the observed optical photometric orbital
period of V1494 Aql (0.13467~d; Retter et al. 2000) within the errors.

\section{Discussion}

There are two obvious and distinct features in these light curves, the
complex rise and fall in the observed count rate, which lasts about
1000 sec, and the periodic signal.  We currently have no explanation
for the burst. The spectrum obtained during the burst is slightly
different from the spectrum of the rest of the observation (Starrfield
et al. 2001).  Once the orbital phase of the burst is known, it may
provide more insight as to the cause.

The periodic signal demands more discussion.   We have done a
period analysis of both  observations and find a 2500 s period in
the light curve.
We also find additional periods present in both the zeroth order
and the dispersed data suggesting that we are not observing the
rotation of the underlying WD. {\it We interpret this
result as the discovery of non-radial g$^+$-mode pulsations in
the hot, rekindled WD.}  At this stage in its evolution
from explosion to quiescence, the hot, luminous WD has a
structure which resembles that of the central stars of planetary
nebulae.  It contains a white dwarf core with a composition of
carbon and oxygen and a thin layer of accreted plus core
material on the surface in which nuclear burning is occurring at
or near the bottom. Both this star and the central stars of
planetary nebulae have effective temperatures exceeding
$10^{5}$K and, when analyzed, the spectrum will show
evidence for hydrogen depletion and a high abundance of carbon
and oxygen.  This star, therefore, resembles the planetary nebulae
nuclei showing O~VI in emission and absorption in optical-UV spectra,
some of which have been shown to pulsate at periods ranging
from $\sim$1000 s to $\sim$5000 s (Ciardullo \& Bond 1996).
The 2500 s period that we have found is close to periods found
in observations of NGC  2371-2, RX J2117+3412, or NGC 1501
for example (Ciardullo \& Bond  1996, and references therein).

Because of the similarity in structure we speculate that pulsation
driving in this star is analogous to that of either the pulsating
Planetary Nebulae Nuclei (PNN) or the GW Vir stars (see
Dreizler et al. 1995 and references therein).  For these stars it
has been proposed that pulsations are driven by the
$\kappa$/$\gamma$ effects---the instability caused by compression-induced 
increases in opacity and ionization---in the partial ionization zones of
carbon and oxygen near the surface (Starrfield et al. 1984,
1985).  In support of the possibility that this mechanism could
be acting in V1494 Aql,  we point out that the nova mechanism
requires CO nuclei to have been mixed up into the accreted
layers and this WD should have enriched CO near the surface.
However, studies at that time suggested that a significant
amount of hydrogen could poison pulsation driving.  Since the
WD is still luminous and presumably still burning hydrogen, it
should still have hydrogen near the surface which would argue
against the $\kappa$/$\gamma$ mechanism.  Recent
observational studies, however, have found pulsations in
members of the GW Vir class which do have hydrogen present
at the surface (Dreizler et al. 1996) and, therefore, the
$\kappa$/$\gamma$ mechanism remains a viable possibility.

If theoretical studies of this star show that the
$\kappa$/$\gamma$ mechanism cannot be responsible for the
oscillations, then another possibility is the $\epsilon$
mechanism.  In this case, pulsation driving is caused by the
ongoing nuclear energy generation near the surface.
The pulsations would then last as long
as there was  nuclear burning in the envelope of the white dwarf.  
However, while
existing studies of this phenomenon either in nova models
(Sastri \& Simon 1973),  high luminosity degenerate stars (Vila
\& Sion 1976), or in hydrogen shell burning PNN (Kawaler
1988) have found instabilities, they are at periods far shorter than we
have observed in V1494 Aql.

\section {Summary}

We have obtained LETG+HRC-S spectra of V1494 Aql about
10 months after discovery.  The X-ray light curve shows a
double peaked, short timescale, rise of $\sim$10 in
counts per second.  We have no explanation for this rise.

Power spectrum analyses of the zeroth order image
shows a dominant period at 2498.8 s.  There are
other periods present our analyses which implies that we are
observing pulsations and not rotation.

A period of $\sim$2500 s falls in the range of those measured
for pulsating PNN such as NGC 1501 or K1-16.  This is to be
expected, if we are observing pulsation, since the structure of the
WD at the time of our observations resembles that of a PNN.

\acknowledgments

We are
grateful to E. M. Sion for valuable scientific discussions and to the
referee who made valuable suggestions for improving the text.  We thank
P.~Ratzlaff for developing the LCURVE software used in this work.  JJD
and YB were supported by NASA contract NAS8-39073 to the {\em Chandra
X-ray Center}.  SS, RMW, PHH, JWT, RDG, and CEW were partially
supported by grants from the Chandra X-ray Center to their various
institutions.  Finally, 
we would like to thank the mission planners and staff of the CXC for
flawless scheduling and execution of this ToO program.





\begin{figure}
\epsscale{0.9}
\plotone{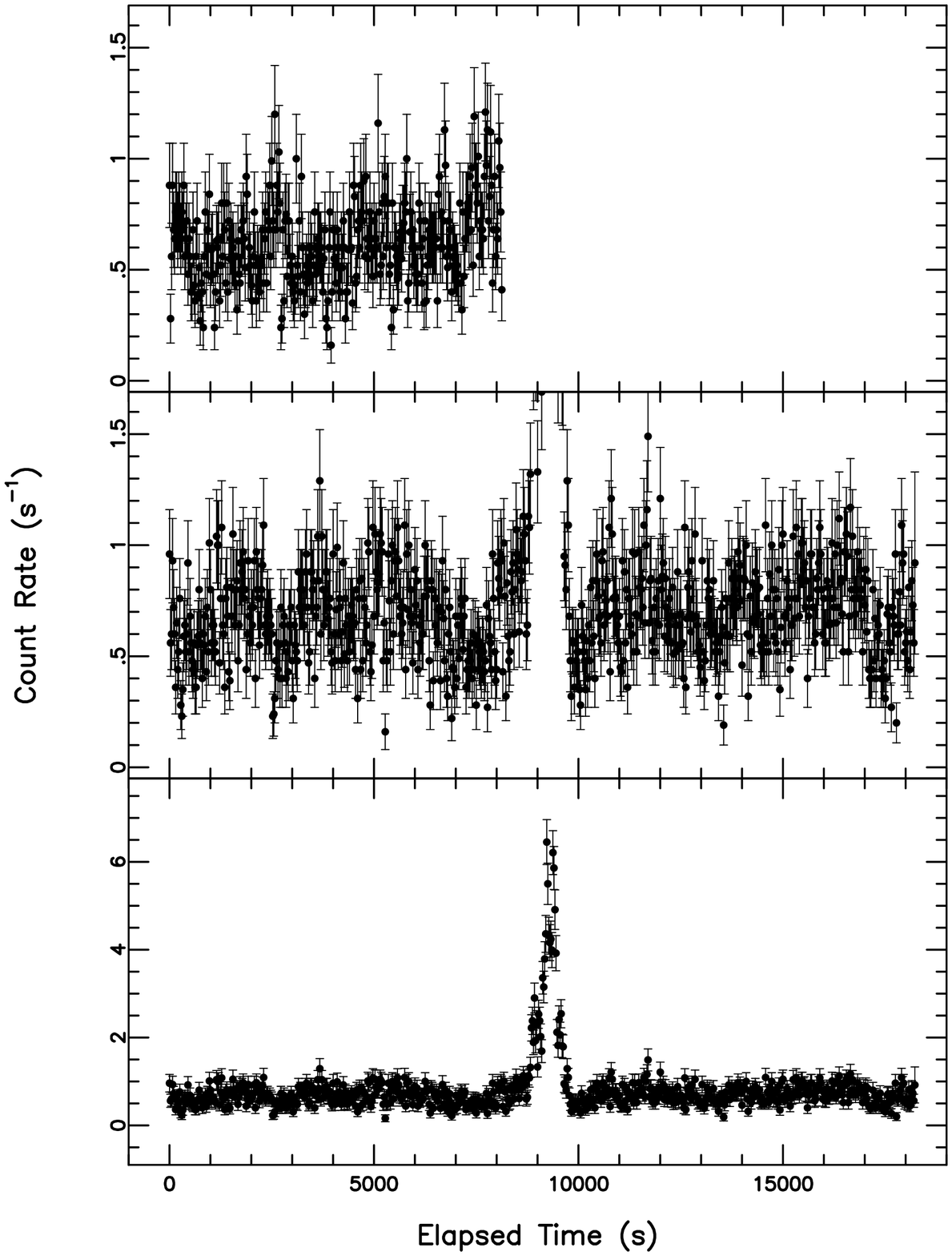}
\caption{{\it Chandra} LETG+HRC-S X-ray light curves of V1494 Aql.
Top: Light curve obtained on 2000 September 28 (ObsID 02308).  Middle:
Light curve obtained on 2000 October 1 (ObsID 00072).  Bottom: Same as
the middle panel but showing the data scaled by the full range of the
count rate to highlight the burst.}
\end{figure}

\begin{figure}
\plotone{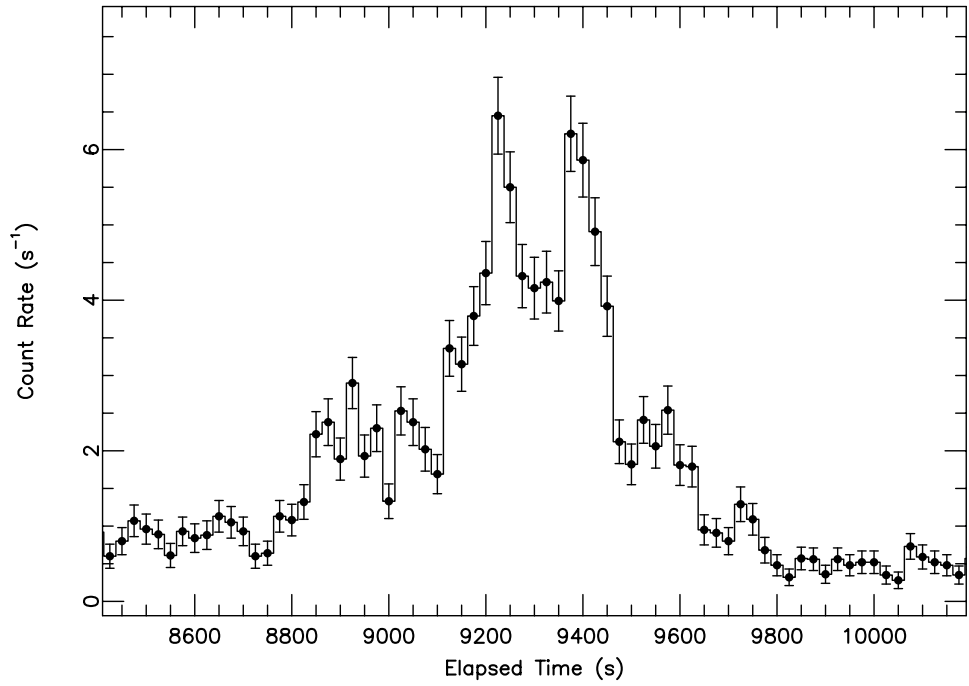}
\caption{Enlargement of the region around the time of the X-ray burst
on 2000 October 1 (ObsID 00072) sampled in 25~s bins.  Note the
presence of two large flares as well as precursor and trailer events.}
\end{figure}

\begin{figure}
\epsscale{0.9}
\plotone{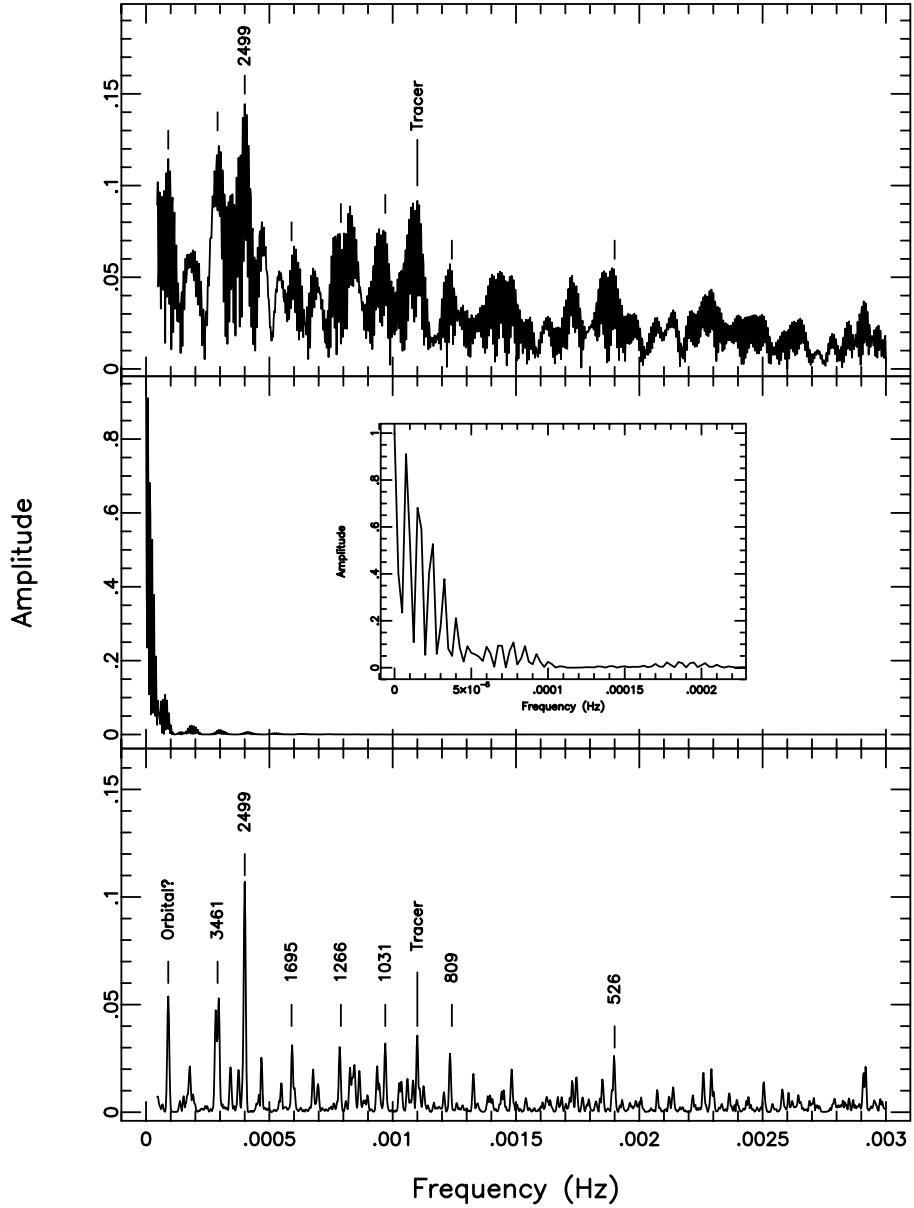}
\caption{Temporal spectra of the X-ray light curves.  Top: Raw or
dirty power spectrum.  Middle: Spectral window function and an
enlargement of the window core (inset).  Bottom: Cleaned power
spectrum.  The significant peaks, the tracer signal with a frequency
of 0.0011~Hz, and a possible orbital (0.13467~d) X-ray modulation are
indicated in both the top and bottom panels.  A period at
$\simeq$2500~s is readily apparent in both dirty and cleaned power
spectra.}
\end{figure}

\end{document}